\documentclass[draft]{agujournal2019}
\usepackage{url} 
\usepackage{lineno}
\usepackage[inline]{trackchanges} 
\usepackage{soul}
\draftfalse

\journalname{Geophysical Research Letters}

\begin{document}

%
%


\title{New results on the direct observations of thermal radio emission from a solar coronal mass ejection}

%
%




\authors{R. Ramesh\affil{1}, A. Kumari\affil{1,3}, C. Kathiravan\affil{1}, 
D. Ketaki\affil{1,2}, T. J. Wang\affil{4,5}}


\affiliation{1}{Indian Institute of Astrophysics, Koramangala 2nd Block, Bangalore, Karnataka, India - 560034}
\affiliation{2}{Department of Physics, Sir Parashurambhau College, Pune, Maharashtra, India - 411 030}
\affiliation{3}{Department of Physics, University of Helsinki, P.O. Box 64, FI-00014 Helsinki, Finland}
\affiliation{4}{Department of Physics, The Catholic University of America, Washington, DC 20064, USA}
\affiliation{5}{NASA Goddard Space Flight Center, Code 671, Greenbelt, MD 20771, USA}





\correspondingauthor{R. Ramesh}{ramesh@iiap.res.in}




\begin{keypoints}
\item We report observations of thermal radio emission from the frontal structure of a CME simultaneously at 80 MHz and 53 MHz.
\item The electron density, mass and magnetic field of the CME were estimated directly from the observed thermal radio emission.
\item Similar observations with larger low frequency radio antenna arrays like LOFAR and SKA are expected to be useful to understand CMEs.
\end{keypoints}

%
%

%
%


\begin{abstract}
We report observations of thermal emission  
from the frontal structure of a coronal mass ejection (CME) using data obtained with the Gauribidanur RAdioheliograPH (GRAPH) simultaneously at 80 MHz and 53 MHz on 2016 May 1. 
The CME was due to activity on the far-side of the Sun, but near its limb. No non-thermal radio burst activity were noticed.
This provided an opportunity to observe the faint thermal radio emission from the CME, and hence directly estimate the electron density, mass, and magnetic field strength of the plasma entrained in the CME. 
Considering that CMEs are mostly observed only in whitelight and reports on their plasma characteristics are also limited, the rare direct radio observations of thermal emission from a CME and independent diagnosis of its plasma parameters   
are important measurements in the field of CME physics. 
\end{abstract}


%
%

%


%
%
%
%

\section{Introduction}

CMEs are large scale and energetic eruptions in the solar atmosphere during which 
$\approx$10$^{12}$-10$^{16}$g of magnetized coronal plasma are ejected into the heliosphere at speeds ranging from $\approx$100-3000km/s  \cite<e.g.,>{Vourlidas2010}.
They are mostly observed in whitelight
using coronagraphs which 
use an occulter to block the 
bright light from the solar photosphere so that structures like CMEs can be observed with better contrast.
But the size of the coronagraph occulters till date has always been larger than that of the photosphere. For example, in the  
\textit{Large Angle and Spectrometric Coronagraph C2} \cite<LASCO C2;>{Brueckner1995} 
on board the \textit{Solar and Heliospheric Observatory} (SOHO), 
the occulter covers a heliocentric distance ($r$) of $\rm{\approx}2.2R_{\odot}$, where $\rm R_{\odot}$
is the radius of the photosphere. 
This prevents observations of the corona present immediately off the solar limb,
in addition to the corona above the solar disk.
Radio observations are unique in this connection since there is no occulter. 
The radio emission associated with and/or from the CMEs can be divided into two classes, thermal and non-thermal \cite<e.g.,>{Vourlidas2004}.
In the non-thermal case, type IV radio bursts due to gyrosynchrotron and/or plasma emission from the electrons in the CME
\cite{Stewart1974,
Wagner1981,Gary1985,Gopalswamy1987,Gopalswamy1989,Gopalswamy1990,Bastian2001,Maia2007,Ramesh2013,
Tun2013,Bain2014,
Sasikumar2014,Hariharan2016a,Carley2017,Morosan2019,Vasanth2019,Mondal2020}, 
type II radio bursts due to plasma emission from the electrons accelerated 
by 
MHD shocks driven by the CME
\cite{Stewart1974,Mann1995,Aurass1997,Gopalswamy2006,Ramesh2010b,Ramesh2012a,
Anshu2017a,Anshu2017b,Chrysaphi2018,Anshu2019,Maguire2020}, 
and type I noise storm continuum due to plasma emission from changes in the coronal magnetic field during a CME 
\cite{Kerdraon1983,Ramesh2000a,Kathiravan2007}
have been widely reported. Compared to this, there are only a few reports of direct detection of CMEs at radio frequencies via 
thermal bremsstrahlung emission  \cite{Sheridan1978,Gopalswamy1992,Gopalswamy1993,Kathiravan2002,Kathiravan2004,
Kathiravan2005,Ramesh2003b,Ramesh2005a}. 
Here we present radio observations of thermal emission from a CME simultaneously at two different frequencies, and compare the nature of the observed emission with published reports.
We were fortunate that there were no flare activity on the visible hemisphere of the Sun whose associated non-thermal radio burst activity would have otherwise probably prevented us from observing the comparatively weak thermal radio emission from the CME \cite<e.g.,>{Bastian1997}. 
For example, the peak flux density of the radio burst observed on 2006 December 6 
was 
${\approx}10^{6}$sfu (sfu=solar flux unit=$\rm 10^{-22}W/m^{2}/Hz$) in the 1-2 GHz frequency range \cite{Gary2019}. Compared to this the flux density of thermal radio emission from the `quiet' Sun is $\sim$few sfu or even less. This calls for a very large dynamic range to observe the radio burst and the `quiet' Sun simultaneously.

\section{Observations}

The radio observations were carried out on 2016 May 1 using the different facilities 
in Gauribidanur observatory 
\cite<https://www.iiap.res.in/?q=centers/radio;>
{Ramesh2011a,Ramesh2014}.
Two-dimensional images obtained with the \textit{Gauribidanur RAdioheliograPH} \cite<GRAPH;>{Ramesh1998,Ramesh1999a,Ramesh2006b}
at 80 MHz and 53 MHz were used to obtain positional information.
The observations were carried out close to the local meridian transit time of the Sun ($\approx$06:30 UT). 
For radio spectral data, we used observations with the \textit{Gauribidanur LOw-frequency Solar Spectrograph} \cite<GLOSS;>{Ebenezer2001,Ebenezer2007,Kishore2014,Hariharan2016b}, \textit{Gauribidanur RAdio Spectro-Polarimeter} \cite<GRASP;> {Sasi2013a,Kishore2015,Hariharan2015}, 
and e-CALLISTO \cite{Monstein2007,Benz2009}. We also used data obtained with the Gauribidanur Radio Interferometric Polarimeter \cite<GRIP;>{Ramesh2005b,Ramesh2008}). 
The combined use of the aforementioned observations help to understand the radio signatures associated with the corresponding solar activity in a better manner \cite<e.g.,>{Sasikumar2013b}.
Observations 
in EUV at 211{\AA} with the \textit{Atmospheric Imaging Assembly} \cite<AIA;>{Lemen2012} on board the \textit{Solar Dynamics Observatory} (SDO), and in whitelight with the 
COR1 coronoagraph of the \textit{Sun-Earth Connection Coronal and Heliospheric Investigation} \cite<SECCHI;>{Howard2008} on board the \textit{Solar TErrestrial RElationship Observatory} (STEREO) and 
SOHO/LASCO were used to supplement the radio observations.

\begin{figure}
\noindent\includegraphics[width=\textwidth]{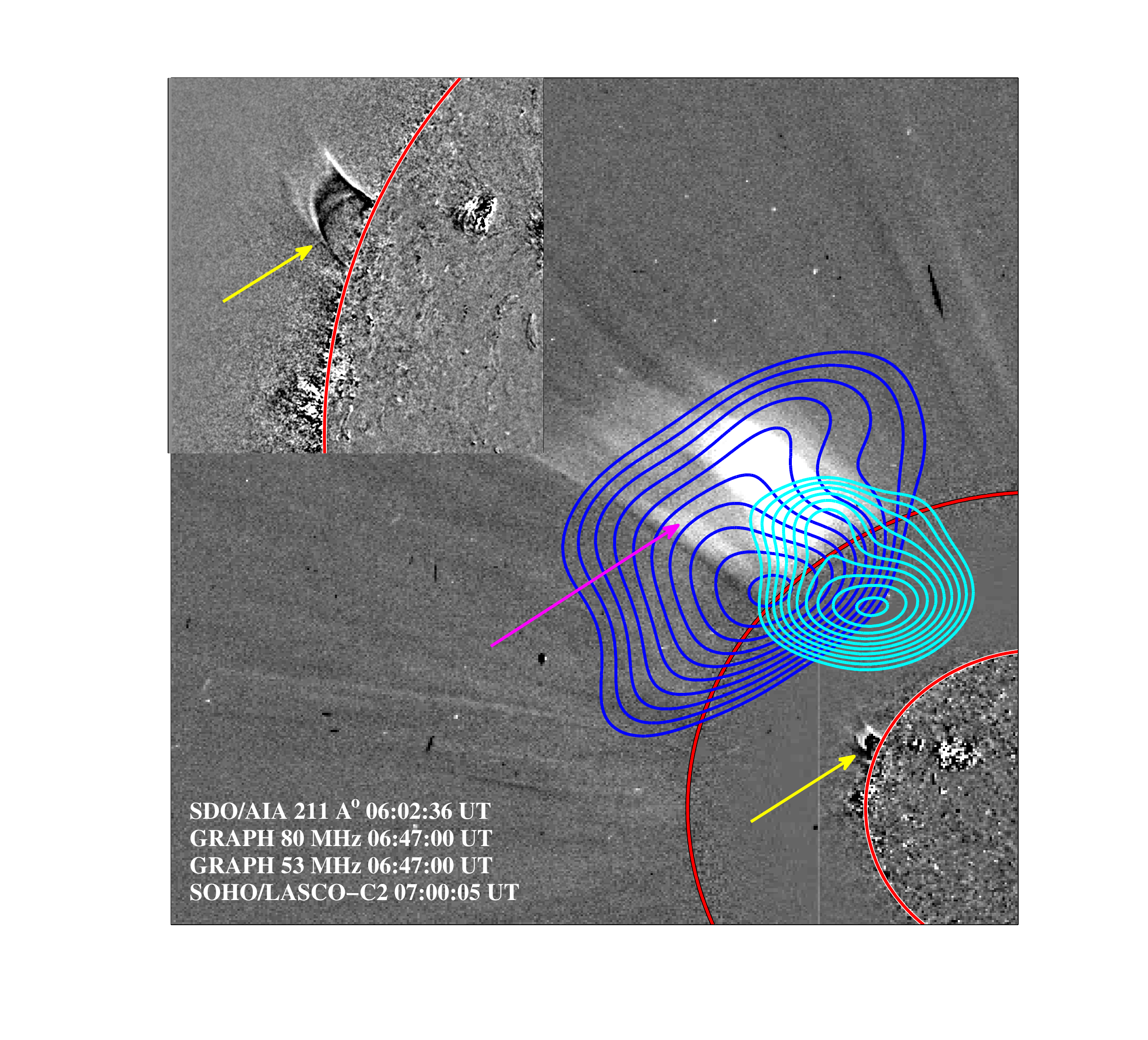}
\caption{A composite of the difference images obtained using EUV 
($\approx$06:02-05:41 UT),
radio ($\approx$06:47-06:42 UT), 
and whitelight ($\approx$07:00-06:00 UT) observations on 2016 May 1.
The inner and outer `red' circles indicate the solar limb (radius$\approx$1$\rm R_{\odot}$), and the occulter in the SOHO/LASCO-C2 coronagraph (radius$\approx$2.2$\rm R_{\odot}$). The bright emission above the occulter (indicated by `magenta' arrow) is the 
whitelight CME. 
The inset in the upper left corner is a close-up view of the region indicated by the `yellow' arrow on the SDO/AIA-211{\AA} image. 
The `cyan' and `blue' contours correspond to radio observations at 80 MHz and 53 MHz. 
The contour levels are ${\approx}$[63, 67, 71, 75, 79, 83, 87, 91, 95, 99]\% of 
$3.4{\times}10^{5}$K (80 MHz) and $0.8{\times}10^{5}$K (53 MHz).}
\label{figure1}
\end{figure}

\begin{figure}
\noindent\includegraphics[width=\textwidth]{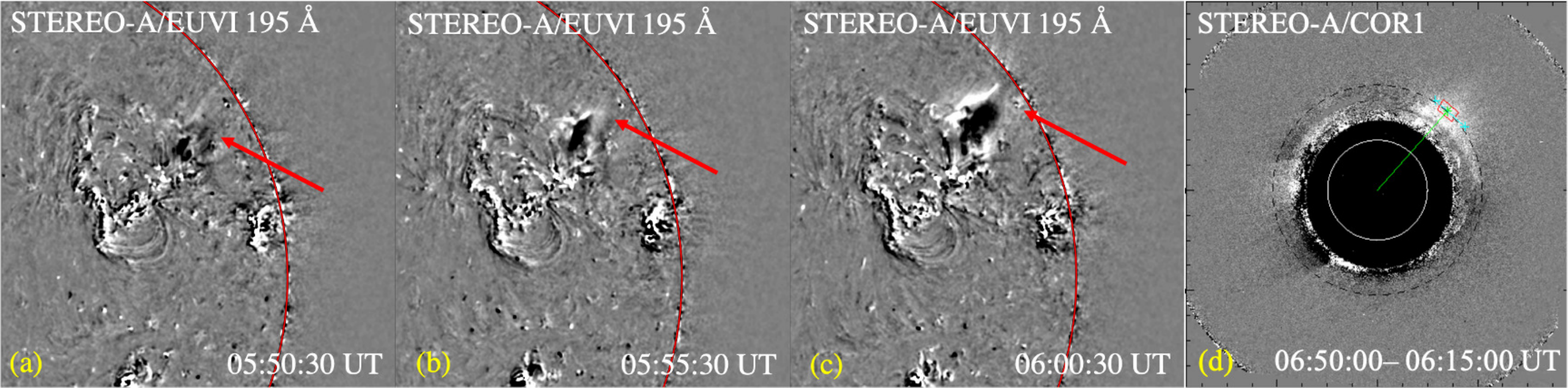}
\caption{Panels (a)-(c) are the STEREO-A/EUVI 195{\AA} running difference images of activity in AR12541 (indicated by `red' arrow) on 2016 May 1. The `red' circle indicate the solar limb. Panel (d) is the STEREO-A/COR1 pB difference image obtained on the same day. The `white' and `black' circles indicate the solar limb
and the coronagraph occulter (radius ${\approx}$1.4$\rm R_{\odot}$), respectively.
The enhanced emission to the upper right of the occulter corresponds to the 
same CME in Figure \ref{figure1}. Its electron density ($N_{e}^{cor1a}$) was estimated from the region covered by the `red' box.}
\label{figure2}
\end{figure}

Figure \ref{figure1} shows a composite of the difference images obtained in EUV, whitelight, and radio on 2016 May 1 during the interval ${\approx}$6-7 UT.
An inspection of the SOHO/LASCO CME catalog (https://cdaw.gsfc.nasa.gov/CME{\_}list/UNIVERSAL/2016{\_}05/univ2016{\_}05.html) indicates that close to the above epoch a CME was observed around position angle (PA, measured counter clockwise from the solar north) $\approx$50$^{\circ}$. The enhanced whitelight emission in Figure \ref{figure1}
corresponds to the frontal structure of the aforementioned CME. 
It was a narrow CME without the three-part structure associated with a typical CME. No streamer-blowout was also noticed (https://lasco{-}www.nrl.navy.mil/carr{\_}maps/c2/).
Its various characteristics estimated from the
SOHO-LASCO observations are, angular width${\approx}$36$^{\circ}$; linear speed in the plane-of-sky${\approx}$482km/s; acceleration${\approx}$-10m/s$^{2}$;
mass${\approx}1.2{\times}10^{15}$g; kinetic energy${\approx}1.4{\times}10^{30}$erg.
An inspection of the SDO/AIA-211{\AA} data indicates that the CME was most likely due to activity in the sunspot region AR12541
located at $\approx$N01E94 (https://www.lmsal.com/solarsoft/ssw/last{\_}events{-}2016/last{\_}events{\_}20160503{\_}1121/index.html).
Since the region was just behind the limb, any projection effects on the aforementioned CME parameters will be very minimal. 
We verified the activity in AR12541 using STEREO-A/EUVI 195{\AA} difference images.
STEREO-A was at $\rm{\approx}E160^{\circ}$ during the above 
period (https://stereo-ssc.nascom.nasa.gov/cgi-bin/make{\_}where{\_}gif). 
This implies that AR12541 was located at 
${\approx}24^{\circ}$ inside the limb for STEREO-A view (see Panels a-c in Figure \ref{figure2}). 

The GRAPH observations in Figure \ref{figure1} were close to $\approx$06:47 UT at 53 MHz and 80 MHz simultaneously. 
The spatio-temporal correspondence between the radio contours and the whitelight CME indicates that the radio emission could be due to the CME. 
The larger size of the radio contours, particularly at 53 MHz and in 
the north-south direction, are likely due to the 
comparatively limited angular resolution of GRAPH
(${\approx}4^{\prime}{\times}6^{\prime}$ (R.A.{$\times$}decl.)
and ${\approx}6^{\prime}{\times}9^{\prime}$
at 80 MHz and 53 MHz, respectively.)
The EUV eruptive activity noticed near the solar limb
indicate the source region and the early phase of the CME. Though it appears that the EUV eruption is slightly displaced from the whitelight and radio structures, an inspection of the 12 sec running difference images using SDO/AIA-211{\AA} and SDO/AIA-193{\AA} indicate that the eruptive activity rises up non-radially in the direction towards the locations of the whitelight and radio structures at a later time. 

The centroids of the radio emission in Figure \ref{figure1} are located at
$r_{80}{\approx}$1.7$\pm 0.2\rm R_{\odot}$ (80 MHz) and 
$r_{53}{\approx}$2.1$\pm0.2\rm R_{\odot}$ (53 MHz). 
Any possible error in the position of the centroids due to propagation effects such as scattering by density inhomogeneities in the solar corona and/or refraction in the Earth's ionosphere is expected to be within the above error limit \cite{Stewart1982,Ramesh1999b,Ramesh2012b,Kathiravan2011,
Mugundhan2016,Mugundhan2018a}. The fact that the Sun is
presently going through a period of extended minimum (during which the observations reported in the present work were carried out) indicates that
the effects of scattering are likely to be less pronounced \cite{Sasikumar2016,Mugundhan2017,Ramesh2020}.
Furthermore, the observations were carried out close to the local noon during which time the zenith angle of the Sun is the least. Possible positional shifts in hour angle and declination due to ionospheric effects are expected to be negligible during that time. Note that the elevation of Sun on 2016 May 1 when the present observations were carried out was $\approx$88$^{\circ}$. 
Secondly, the 53 MHz and 80 MHz radio images in Figure \ref{figure1} were obtained by subtracting the corresponding radio images obtained $\approx$5min earlier at the respective frequencies. 
This time interval is lesser than the period 
($\approx$20min) over which radio source positions at low frequencies usually 
change due to ionospheric effects \cite{Stewart1982,Mercier1996}. 
The brightness temperatures ($T_{b}$) of the radio sources 
in Figure \ref{figure1} near their centroids are $T_{b}^{80}{\approx}3.4{\times}10^{5}$K (80 MHz) and $T_{b}^{53}{\approx}0.8{\times}10^{5}$K (53 MHz).
The corresponding flux density ($S$) values are $S_{80}{\approx}0.14$sfu (80 MHz) and
$S_{53}{\approx}0.06$sfu (53 MHz). The spectral index ($\alpha$) derived from the above flux densities is ${\approx}2.1.$
These are in agreement with that reported for thermal emission from the solar corona at low frequencies \cite{Erickson1977,Sheridan1985,KRS1988,Ramesh2000b,Ramesh2006a,Ramesh2010a}. 
Furthermore, no 
non-thermal radio bursts were observed either elsewhere (ftp://ftp.swpc.noaa.gov/pub/warehouse) or with GLOSS or GRASP during the particular observing period. 
Therefore it is likely that the enhanced radio emission in Figure \ref{figure1} is thermal in nature. The contrast of such emission is generally better at frequencies $<$100 MHz 
\cite<e.g.,>{Lantos1987}. 
Model calculations by \citeA{Bastian1997} also indicate that it is possible to observe enhanced thermal emission from particularly the off-limb CMEs using difference images as in the present case. The above $T_{b}$ values are reasonably consistent with those predicted by the aforementioned model for thermal emission from a `typical' CME in the same frequency range. Note that for non-thermal gyrosynchrotron emission from a CME, the $T_{b}$ values predicted by the model are higher than the aforementioned observed $T_{b}$ values by about factor of four, even for the 
lowest values assumed for the different parameters in the model.
Low frequency observations reported earlier also indicate that the observed $T_{b}$ of gyro-synchrotron emission from a CME are ${\sim}10^{7}{-}10^{8}$K \cite{Wagner1981,Gary1985,Dulk1985,Gopalswamy1987,Gopalswamy1990,Sasikumar2014}.

\begin{figure}
\noindent\includegraphics[width=\textwidth]{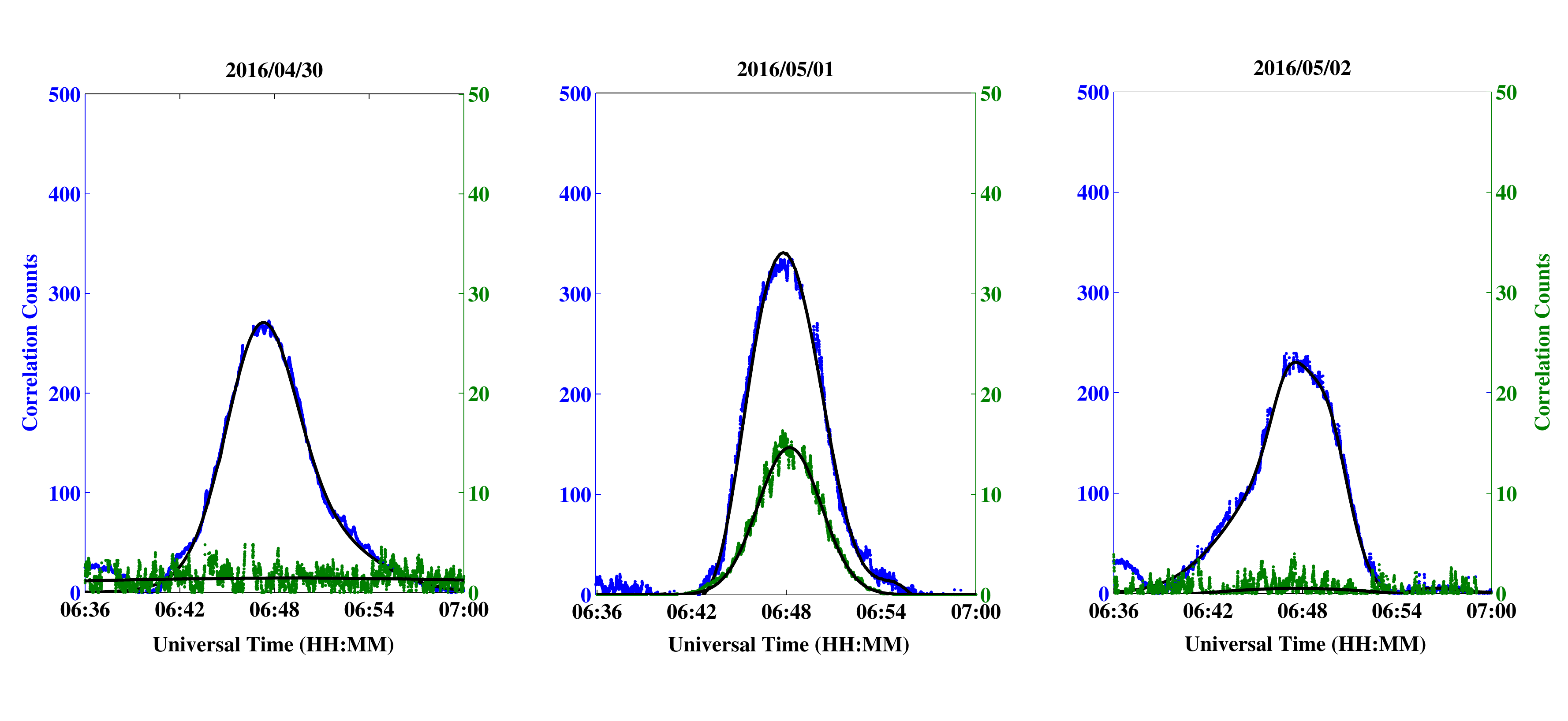}
\caption{GRIP observations of Stokes I (`blue') and V (`green') emission at 80 MHz 
in the transit mode around 06:48 UT. 
The assymetry in the Stokes I observations on 2016 May 2 was due to a local radio frequency interference (RFI) around ${\approx}$06:40 UT. 
The `solid' lines in black colour are the `fit' to the respective data points.}
\label{figure3}
\end{figure}

Figure \ref{figure3} shows the GRIP observations in the transit mode at 80 MHz on 2016 April 30, May 1, and May 2 around ${\approx}$06:48 UT. The half-power width of the response pattern of GRIP at the above frequency is 
${\approx}1.8^{\circ}$. 
Since this is wider than the size of the Sun,
the observations in Figure \ref{figure3} correspond to integrated emission from the `whole' Sun at 80 MHz. The gradual increase/decrease in the observed correlation count is due to the passage of Sun across the response pattern of  GRIP.
There is noticeable Stokes V emission on 2016 May 1 with a peak correlation count of {$\approx$}15. No similar emission was observed on 2016 April 30 and May 2. The Stokes I emission on 2016 May 1 also shows
an increase of ${\approx}$125 counts
as compared to the observations
on 2016 May 2 (i.e. from ${\approx}$225 to ${\approx}$350 counts). The Sun was `quiet' on 2016 April 30 too, but the peak Stokes I correlation count 
(${\approx}$255) was slightly higher.
Following \citeA{Kundu1977}, we used 2016 May 2 observations as the reference value for `quiet' Sun to estimate the aforementioned increase in the Stokes I correlation count on 2016 May 1.
Based on the above, we calculated the degree of circular polarization
($dcp{=}\frac{|V|}{I}$) for the observations on 2016 May 1, and the value is 
$\approx$12$\%$. This is lesser than the $dcp$ reported for non-thermal gyrosynchrotron radio emission from CME associated radio emission in the similar frequency range \cite{Dulk1985,Sasikumar2014}.
The absence of radio bursts during our observing period on 2016 May 1 argue against any non-thermal origin for the observed Stokes V emission \cite<e.g.,>{Mugundhan2018b}. The possibility of non-thermal noise storm continuum emission too can be ruled out since the sunspot region AR12541 in which there was activity that day was behind the limb as mentioned earlier and noise storm sources are highly directive \cite{Ramesh2011b}.
Coincidentally, the GRIP observations on 2016 May 1 was nearly at the same time as the GRAPH observations in Figure \ref{figure1}. One of the observing frequencies with the GRAPH (80 MHz) was also same as that of the GRIP. Therefore it is possible that the observed Stokes V emission on 2016 May 1 in Figure \ref{figure3} could be due to the enhanced thermal emission above the limb in Figure \ref{figure1} at the same frequency \cite{Sastry2009,Ramesh2010a}.
Note that the thermal emision could be circularly polarized in the presence of a magnetic field.
Since the medium becomes birefringent due to the latter,
the randomly polarized thermal radiation propagates in two orthogonal circular modes, i.e., the ordinary (`o') and the extraordinary (`e') mode. A difference between the optical depths and hence the $T_{b}$ of the two modes will result in a finite  value of $dcp$, if the corona is not optically thick.

\section{Analysis and Results}

\begin{figure}
\noindent\includegraphics[width=\textwidth]{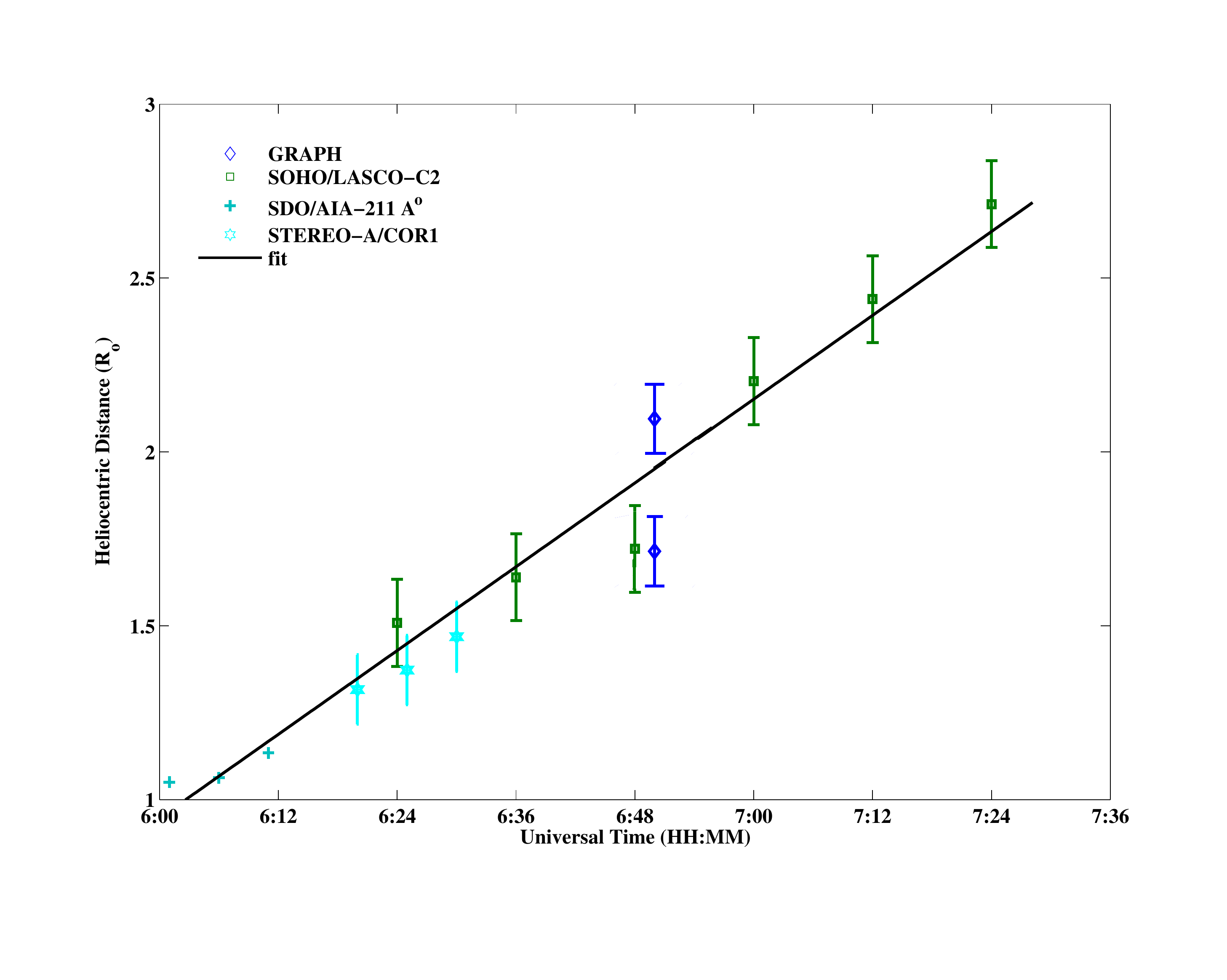}
\caption{CME h-t plot obtained using EUV, whitelight and radio observations.}
\label{figure4}
\end{figure}

According to SOHO/LASCO CME catalog, the LE of whitelight CME in Figure \ref{figure1} was at $r_{cme}{\approx}$3.86$\rm R_{\odot}$ around 
$\approx$06:48 UT, nearly the same epoch as the GRAPH observations. But the centroids
of the radio emission are located at   
$r_{80}{\approx}$1.7$\pm0.2\rm R_{\odot}$ (80 MHz) and 
$r_{53}{\approx}$2.1$\pm0.2\rm R_{\odot}$ (53 MHz). Therefore it is likely that
the radio sources are not associated with the LE of the frontal structure in the whitelight CME. They seem to correlate spatially with parts of the frontal structure behind its LE. 
The separation between the centroids of the 80 MHz and 53 MHz radio sources in Figure \ref{figure1} suggests possible density gradient in the frontal structure similar to that in the background corona since it is well known that radio emission from the Sun at a particular frequency ($f$) can propagate towards the observer only from/above the critical level at which the plasma frequency ($f_{p}$) corresponding to the local electron density ($N_{e}$) equals $f$.
Note that a gradually increasing $N_{e}$ (towards the Sunward side) in the CME frontal structure is possible if there is material pile-up from lower coronal levels \cite{Subramanian2007,Bein2013}. 
We generated a h-t plot using centroids of EUV eruptive activity, frontal structure of whitelight CME, and radio sources (Figure \ref{figure4}).
To identify the centroids of EUV and whitelight structures, we fitted each of them with a circumcircle. The LE of the corresponding structure and the solar limb were assumed to be the upper and lower boundaries \cite<e.g.,>{Bein2013}. 
The h-t measurements obtained with STEREO-A/COR1 were multiplied by 1/cos($24^{\circ}$) to correct for projection effects. All the data points are reasonably well aligned with the linear least squares fit line. 
The speed of the CME ($v_{cme}$) estimated from the aforesaid fit is 
$\approx$233km/s. Note that since GRAPH observations were confined to 80 MHz and 53 MHz, there are only two data points from GRAPH.

It is known that the larger density gradient near the solar limb leads to refraction of low frequency radio waves. So, the contribution to observed thermal 
bremsstrahlung emission at any given frequency is primarily from regions well above the corresponding plasma level. This implies smaller optical depth ($\tau$) since the absorption coefficient and hence $\tau$ are maximum only near the plasma level
\cite<e.g.,>{Smerd1950,Aubier1971,Vocks2018}.
In the case of solar corona, $T_{b}$ and $\tau$ are related to its electron temperature ($T_{e}{\approx}10^{6}$K) as 
$T_{b}{\approx}T_{e}(1{-}e^{-\tau})$. 
Since $\tau$ above the limb is 
small as mentioned,  
$T_{b}{<}T_{e}$ 
although the spectral index could be indicative of optically thick thermal emission.
The emission from CMEs too undergo
refraction and reflection near the plasma layer as described above \cite<e.g.,>{Bastian1997}. The fact that $T_{b}^{80}$ and 
$T_{b}^{53}$ in the present case (see Section 2) are lesser than $T_{e}$ 
also indicates the same. Note that since a CME contains primarily  coronal material, its $T_{e}$ is considered to be the same as that of the surrounding corona \cite<e.g.,>{Vourlidas2004}. 
Using the aforementioned relation, we calculated $\tau$ corresponding to 
$T_{b}^{80}$ and $T_{b}^{53}$, and the values are 
${\tau}_{80}{\approx}$0.4 and ${\tau}_{53}{\approx}$0.1. These are reasonably consistent with $\tau$ values for similar frequencies and at locations above the solar limb as in the present work
\cite<e.g.,>{Smerd1950,Thejappa1992,Vocks2018}. Based on this, we then estimated $N_{e}$ at $r_{80}{\approx}$1.7$\pm0.2\rm R_{\odot}$ and 
$r_{53}{\approx}$2.1$\pm0.2\rm R_{\odot}$ 
using the relation 
${\tau}{=}2{\times}10^{-24}N_{e}^{2}LT_{e}^{-1.5}f^{-2}{\eta}^{-1}$,
where 
$L$ (cm) is the 
width of radio source along the line of sight and
${\eta}{=}\sqrt{1{-}80.6{\times}10^{6}N_{e}f^{-2}}$ is the refractive index 
\cite<e.g.,>{Sheridan1985}.
The lateral and radial widths 
of the radio sources in Figure \ref{figure1} are 
$w_{80}{\approx}15^{\prime}{\times}16^{\prime}$ (80 MHz) and 
$w_{53}{\approx}26^{\prime}{\times}29^{\prime}$ (53 MHz).
Assuming $L$ as the average of lateral and radial widths, 
we find that 
$N_{e}^{80}{\approx}$2.2$\rm{\times}10^{7}cm^{-3}$ at 
$r_{80}{\approx}$1.7$\pm$0.2$\rm R_{\odot}$ and 
$N_{e}^{53}{\approx}$1.0$\rm{\times}10^{6}cm^{-3}$ at
$r_{53}{\approx}$2.1$\pm$0.2$\rm R_{\odot}$.
These are in good agreement with our independent estimates of CME density using  linearly polarized brightness (pB) measurements with STEREO-A/COR1 and the inversion technique based on spherically symmetric 
polynomial approximation (SSPA; solar.physics.montana.edu/wangtj/sspa.tar; \citeA{Wang2014,Wang2017}). 
At $\approx$06:50 UT (about the same time as GRAPH observations), we find
$N_{e}^{cor1a}{\approx}$4.8${\pm}$1.4$\rm{\times}10^{6}cm^{-3}$ 
near $r{\approx}$2.1$\rm R_{\odot}$ (see Figure \ref{figure2}).

Considering the coronal plasma as fully ionized gas of normal solar composition (90\% hydrogen and 10\% helium by number), one finds that each electron is associated with approximately $2{\times}10^{-24}$g of material. Therefore the mass of the enhancement is given by
$M{=}2{\times}10^{-24}N_{e}V$,
where $V$ is the volume of enhancement. We calculated $V$ as $w_{80}{\times}L_{80}$ and $w_{53}{\times}L_{53}$ at 80 MHz and 53 MHz, respectively.
For $N_{e}$ at the above two frequencies, we used $N_{e}^{80}$ and $N_{e}^{53}$ calculated earlier.
Substituting for different values in the above relation, we get 
$M_{80}{\approx}1.3{\times}10^{16}$g (80 MHz) and 
$M_{53}{\approx}3.4{\times}10^{15}$g (53 MHz).
The average of the above values is $M_{cme}^{radio}{\approx}7.4{\times}10^{15}$g. 
Using this and $v_{cme}$ mentioned earlier, we calculated the CME kinetic energy  
$E_{cme}^{radio}{\approx}2{\times}10^{30}$erg.
The estimates of $M_{cme}^{radio}$ and $E_{cme}^{radio}$ 
agree reasonably with the
mass and kinetic energy of the CME 
estimated using SOHO/LASCO-C2 observations (see Section 2). 
Having said so, we would like to mention here that the values of $M_{cme}^{radio}$ and $E_{cme}^{radio}$ calculated above should be considered as upper limits due to lack of information on $L$.
For example, if we assume that $L$
is equal to density scale height in the corona 
(i.e. ${\approx}10^{5}$km), 
then both $M_{cme}^{radio}$ and 
$E_{cme}^{radio}$ should be lesser by about an order of magnitude. 
So, the more appropriate results 
would be 
$M_{cme}^{radio}{\sim}10^{15}$g and $E_{cme}^{radio}{\sim}10^{30}$erg. Note that  estimates of CME mass using whitelight coronagraph observations too are affected by uncertainities \cite{Vourlidas2000,Carley2012,Bein2013}.
Moving to magnetic field strength ($B$) calculations, \citeA{Sastry2009} had shown earlier that if 
$B{\approx}$0.5G and
electron density is
${\approx}1.2{\times}10^{6}\rm cm^{-3}$, then it should be possible to observe circular polarized radio emission with $dcp{\approx}$12\%.
In the present case, 
the density at $r{\approx}$2.1$\rm R_{\odot}$ is 
${\approx}1.0{\times}10^{6}\rm cm^{-3}$
and the estimated $dcp$ due to CME associated enhanced radio emission is 
${\approx}12\%$ as mentioned earlier. Therefore it is likely that magnetic field strength in the CME at the above location is ${\approx}0.5$G. This is close to some of the earlier reported $B$ values at nearly the same heliocentric distance from other type of observations like non-thermal gyrosynchrotron emission from the CME loops \cite{Bastian2001}, geometrical properties of the CME flux rope and the shock at its leading edge \cite{Gopalswamy2012}, CME associated fast magnetosonic waves \cite{Kwon2013}, and second harmonic plasma emission from the CME leading edge \cite{Hariharan2016a}. Model calculations reported by \citeA{Zucca2014} in connection with a CME associated coronal type II burst also predict nearly the same value of $B$ at the above $r$. 

\section{Summary}

We have presented 
evidence for enhanced thermal 
radio emission associated with the frontal structure of a CME (from just behind the limb of the Sun)
that was observed with SOHO/LASCO-C2 on 2016 May 1. 
The radio data were obtained with GRAPH at 80 MHz and 53 MHz simultaneously. 
The plasma characteristics of the CME, estimated 
from radio observations, are:
$N_{e}{\approx}2.2\rm{\times}10^{7}cm^{-3}$ at 
$r{\approx}1.7\rm R_{\odot}$ and 
$N_{e}{\approx}1.0\rm{\times}10^{6}cm^{-3}$ at 
$r{\approx}2.1\rm R_{\odot}$; 
mass and kinetic energy are ${\sim}10^{15}$g and 
${\sim}10^{30}$erg, respectively; magnetic field strength is
$B{\approx}$0.5G at $r{\approx}2.1\rm R_{\odot}$.
Future observations with low frequency radio antenna arrays with larger collecting area like LOFAR and SKA should be able to effectively exploit this possibility for CMEs against the solar disk also \cite<e.g.,>{Ramesh2000c}.

\acknowledgments
We thank the staff of Gauribidanur observatory for their
help in maintenance of antenna, receiver systems and the observations.
Ajith Sampath and S. Kokila are thanked for their contributions to the work. 
The SOHO-LASCO CME catalog is generated and maintained at CDAW Data Center
by NASA and Catholic University of America in cooperation with 
Naval Research Laboratory. 
The SDO/AIA data are courtesy of NASA/SDO and AIA science teams. 
The work of TJW was supported by NASA Cooperative Agreement NNG11PL10A to CUA and NASA grants 80NSSC18K1131 and 80NSSC18K0668. Data used in the study are at https://www.iiap.res.in/gauribidanur/home.html. We thank the
referee for his/her comments which helped us to present the results in a better manner.


%
%


%
%
%
%
%

\end{document}